\documentclass{pasj00}
\draft

\begin{document}
\SetRunningHead{Kotnik-Karuza et al.}{RR Tel non-Fe$^+$ emission
lines }
\Received{2000/12/31}
\Accepted{2001/01/01}

\title{Variation of fluxes of RR Tel emission lines measured in 2000 with respect to 1996}

\author{Dubravka \textsc{Kotnik-Karuza} %
  \thanks{ Present address: Department of Physics, University of Rijeka, Omladinska
14, HR-51000 Rijeka, Croatia}} \affil{Department of Physics,
University of Rijeka, Omladinska 14, HR-51000 Rijeka, Croatia}
\email{kotnik@ffri.hr}

\author{Michael \textsc{Friedjung}}
\affil{Institut d'Astrophysique de Paris UMR7075 CNRS, Universit\'e
Pierre \& Marie Curie, 98 bis Boulevard Arago, 75014 Paris,
France}\email{fried@iap.fr} \and
\author{Katrina {\sc Exter}}
\affil{Instituto de Astrofisica de Canarias, La Laguna (Tenerife),
Espana}\email{katrina@iac.es}

%


 \maketitle

\begin{abstract}
The aim of this work is to make available unpublished non-Fe$^+$
emission line fluxes from optical spectra of the symbiotic nova RR
Tel which were taken in 2000, and to compare them with fluxes of the
same lines from spectra taken in 1996. After leaving out blends and
misidentifications, as well as the unreliable far-red and violet
lines, we present the log\,(F$_{2000}/$F$_{1996}$) flux ratios for
identified non-Fe$^+$ lines. Mean values of
log\,(F$_{2000}/$F$_{1996}$) for different ionization
   potential ranges of the ions producing the lines are shown separately for
   the permitted and forbidden lines. All means show fading, which is larger
   in the lowest range of ionization potential. Provisional interpretations are
   suggested. We also measured the values of FWHM in 2000;
   the previously known decrease with time of FWHM of lines due to the same ion has
   continued.

\end{abstract}

\section{Introduction}

\
\\
Symbiotic binaries consist of a cool giant and a more compact
companion, which is thought to be in most cases a white dwarf.
Accretion occurs from the cool giant to its compact companion.
States of ``activity'' shown by brightening of the compact companion
and spectral changes are now believed to be at least often
associated with thermonuclear burning in the outer layers of the
white dwarf. Some symbiotic binaries are called ``symbiotic novae'',
as only one activity stage, associated with a larger amplitude
brightening than the brightenings of other symbiotic binaries, has
been observed. In a subclass of symbiotic binaries called symbiotic
Miras the cool giant is a Mira variable and is surrounded by dust
(\cite {W87}). Accretion in symbiotic Miras is thought to occur from
the strong wind of the giant. This sort of symbiotic binary unlike
the large majority of normal single Miras undergoes obscuration
events, involving an increase of absorption by dust (\cite {W03}).

The compact component of RR Tel is both a symbiotic nova, with an
outburst in 1944,  and a symbiotic Mira binary system. It has faded
in the optical since 1950, with its spectrum showing emission lines
from atoms whose maximum ionization potential increased with time
over many years, while the lines from more ionized atoms tended to
be wider. This development was summarized by Thackeray (\cite
{T77}). Ultraviolet observations with IUE were described by Penston
et al (\cite{Pe83}) In addition, like other symbiotic Miras, RR Tel
has dust obscuration events, which produce decreasing continuum
fluxes in the infrared and decreasing optical Fe$^+$ emission line
fluxes. In previous work we studied time variable and wavelength
dependent absorption of the continuum of the cool component in the
infrared, the characteristics of the regions producing permitted and
forbidden Fe$^+$ emission lines, and the variation of the Fe$^+$
emission line fluxes due to variable obscuration and what may be
considered a dust obscuration event at the time of observations in
2000 ( \cite{KK02};  \cite{KF06}). The variation Fe$^+$ emission
line flux, between an epoch of much dust absorption and one of less
dust absorption was relatively simple and might be interpretable as
due to the presence of separate, optically thick, absorbing dust
clouds further from the cool component than the wavelength dependent
infrared absorption.
\\

\newpage

\section{Observations and methods of analysis}

Optical spectra of RR Tel taken with the Anglo Australian Telescope
in 1996 and 2000 have been compared. The former was obtained and
calibrated by  Crawford et al ( \cite{Cr99}) with a resolution of
about 50000 . The second spectrum, taken in July 2000 in the
wavelength range of 3180-8000 \AA\ with almost twice the spectral
resolution, was flux calibrated with an HST spectrum taken in
October 2000. The values of line flux $F_{2000}$ were dereddened
using R = 3.1, the Howarth reddening law ( \cite{Ho83}) and E(B-V)=
0.08 ( \cite{Jo94}). The line fluxes were measured by Gaussian
fitting. Crawford et al corrected the 1996 spectra for interstellar
extinction using the coefficients listed in Cardelli et al (
\cite{Ca89}) and the same E(B-V)= 0.08.

We identified and measured 523 emission lines, covering the
wavelength region 3180 - 9230 \AA, most of which were found in the
spectra of RR Tel observed in 1996 ( \cite{Cr99}). In order to
compare lines from the two spectra, ratios of fluxes were determined
for each line. The flux ratios in the far red ($\lambda$ $>$ 7000
\AA) and far violet ($\lambda$ $<$ 3400 \AA) have been left out from
our study because of greater measurement errors in these regions.
After elimination of blends and misidentifications a set of 490
lines  was accepted for further analysis. For each line in Table 1 (
available in electronic form at the ....) the following information
has been given: laboratory wavelength, ion identification, multiplet
number, ionization potential of the contributing ion, line flux
$F_{2000}$ (in units of 10$^{-12}$erg s$^{-1}$cm$^{-2}$) and FWHM
(km s$^{-1}$)from the spectrum taken in 2000, line flux $F_{1996}$
from the spectrum taken in 1996 in the same flux units, excitation
mechanism and statistical weight as a measure of the error of the
flux determination.  Note that the multiplet number is missing for
lines for which this labelling has not been given in the databases
used for identification. The missing FWHM values refer to lines that
could not be satisfactorily Gaussian fitted and their flux was
obtained by adding the counts above a defined continuum region over
a defined lambda range . The mean FWHM values for each ion derived
from the corresponding values of individual lines together with the
standard deviations of the means are shown in Table~\ref{tab:dva}.
 In this paper we analyze 367 lines of ions other than FeII, while
the remaining 156 Fe$^{+}$ lines were the subject of our previous
paper ( \cite{KF06}).

Errors were calculated by performing multiple  measurements of the
lines, as well as comparing the fluxes measured from overlapping
orders and overlapping gratings. The error in each line flux over
the wavelength range 4920 $<$ $\lambda$ $<$ 6370 \AA\   was
estimated as up to 25\%.  Outside this wavelength range, except for
the very blue lines, the error is about 15\%.  For the bright lines
the errors lie considerably under these limits, while for the faint
lines,
 due to the uncertainty in the flux calibration process, the errors can exceed the given limits
 even by 15 \%.
 For $\lambda$ $>$ 7000 \AA\  not included in our analysis,
 where the calibration had  to be extrapolated beyond the HST spectrum,
 the calibration and measurement uncertainties yield an average error of 28\%.

We took account of the errors in line fluxes by separating the lines
into three groups and by assigning them different statistical
weights. In our notation, the group "a" has the largest weight 3 and
contains lines with errors smaller than 15\%. Weak lines and lines
from the less certain wavelength range are in group "b" which has a
weight of 2. In the group "c" with the smallest weight of 1 the very
weak lines, those identified with a certain degree of uncertainty,
badly fitted lines and possible blends are included.

\newpage
\setcounter{table}{1}
\begin{longtable}{*{10}{l}}
\caption{Mean FWHM values for individual ions from the spectra of RR
Tel taken in 2000} \label{tab:dva}\hline \hline
Ion&  FWHM & N & ${\sigma}$ & IP & Ion&  FWHM & N & ${\sigma}$ & IP \\
   &(km/s)& & & (eV) &  &(km/s)& & & (eV)\\
\hline
\endhead
\hline
\endfoot
\hline \multicolumn{10}{l}{B refers to the lines excited by Bowen
mechanism while the nB lines are excited by recombination.

} \\
\hline
\endlastfoot

HI  &   64.0    &   24  &   5.1 &   13.598  &   [ArIV]  &   68.0    &   1   &       &   40.74   \\
HeI &   40.4    &   24  &   2.4 &   24.587  &   [ArV]   &   50.0    &   2   &   2.7 &   59.81   \\
HeII    &   56.0    &   26  &   1.8 &   54.416  &   ArII    &   24.0    &   1   &       &   27.629  \\
CII &   56.0    &   3   &   22.7    &   11.26   &   [KIV]   &   41.0    &   1   &       &   45.806  \\
CIII    &   45.0    &   7   &   3.7 &   47.888  &   [KV]    &   92.0    &   1   &       &   60.91   \\
CIV &   50.7    &   2   &   3.5 &   64.492  &   [CaV]   &   61.0    &   2   &   8.8 &   67.27   \\
$[$NI]    &   18.0    &   1   &       &   0   &   [CaVI]  &   80.0    &   1   &       &   84.5    \\
$[$NII]   &   105 &   1   &       &   14.534  &   CaI &   66.0    &   3   &       &   0   \\
NI  &   18.0    &   1   &       &   14.534  &   ScI &   13.0    &   2   &   7.2 &   0   \\
NII &   24.7    &   3   &   2.4 &   29.601  &   ScII    &   54.0    &   1   &       &   6.561   \\
NIII    &   33.0    &   10  &   3.7 &   47.448  &   TiI &   45.0    &   3   &   18.4    &   0   \\
NIV &   81.0    &   1   &       &   77.472  &   TiII    &   19.9    &   7   &   5.3 &   6.828   \\
NV  &   59.0    &   1   &       &   97.89   &   [VII]   &   14.9    &   3   &   1.2 &   6.746   \\
$[$OI]    &   15.8    &   2   &   0.0 &   0   &   VII &   25.0    &   10  &   6.5 &   6.746   \\
$[$OIII]  &   55.7    &   3   &   8.4 &   35.117  &   [CrII]  &   11.0    &   2   &   4.0 &   6.766   \\
OI  &   22.0    &   1   &       &   13.618  &   [CrV]   &   72.0    &   1   &       &   49.16   \\
OII &   37.0    &   15  &   7.3 &   35.117  &   CrII    &   44.0    &   10  &   9.8 &   6.766   \\
OIII  B &   35.0    &   7   &   5.8 &   35.117  &   [MnIII] &   115.0   &   1   &       &   15.64   \\
OIII nB &   36.0    &   6   &   4.6 &   54.93   &   [MnIV]  &   24.3    &   3   &   13.0    &   33.668  \\
OIV &   43.0    &   6   &   8.2 &   77.413  &   MnI &   68.0    &   1   &       &   0   \\
$[$FIV]   &   32.0    &   1   &       &   62.708  &   MnII    &   54.0    &   9   &   8.8 &   7.434   \\
FII &   75.0    &   1   &       &   34.97   &   [FeII]  &   28.3    &   58  &   2.9 &   7.902   \\
$[$NeIII] &   35.0    &   3   &   10.6    &   40.963  &   [FeIII] &   51.0    &   13  &   4.6 &   16.188  \\
$[$NeIV]  &   45.2    &   3   &   1.2 &   63.45   &   [FeIV]  &   39.5    &   11  &   2.7 &   30.652  \\
NeII    &   47.8    &   4   &   21.5    &   40.963  &   [FeV]   &   49.2    &   10  &   4.8 &   54.8    \\
$[$NaV]   &   54.0    &   1   &       &   98.91   &   [FeVI]  &   49.5    &   10  &   4.7 &   75  \\
MgI &   22.0    &   1   &       &   0   &   [FeVII] &   69.9    &   3   &   1.1 &   99.1    \\
MgII    &   15.0    &   1   &       &   15.035  &   FeII    &   18.0    &   81  &   1.6 &   7.902   \\
AlII    &   70.0    &   1   &       &   18.828  &   FeIII   &   27.2    &   6   &   5.2 &   16.188  \\
SiI &   34.0    &   1   &       &   0   &   [CoII]  &   29.0    &   1   &       &   7.881   \\
SiII    &   29.5    &   11  &   3.7 &   16.345  &   [CoVII] &   83.0    &   1   &       &   103 \\
SiIV    &   64.0    &   1   &       &   45.142  &   [NiII]  &   27.0    &   4   &   10.1    &   7.639   \\
$[$SII]   &   19.2    &   2   &   2.9 &   10.36   &   [NiIV]  &   26.0    &   1   &       &   35.19   \\
$[$SIII]  &   55.0    &   2   &   5.0 &   23.338  &   [NiV]   &   36.0    &   1   &       &   54.9    \\
SII &   39.6    &   5   &   9.4 &   23.337  &   [NiVIII]    &   50.0    &   1   &       &   134 \\
SIII    &   18.5    &   2   &   0.5 &   34.79   &   NiII    &   38.0    &   1   &       &   7.639   \\
ClII    &   67.0    &   1   &       &   23.814  &   NiIII   &   17.0    &   1   &       &   35.19   \\
$[$ArIII] &   41.0    &   1   &       &   27.629  &   NiV &   20.0    &   1   &       &   75  \\

\end{longtable}

\newpage

\section{Data analysis}

\begin{figure}
  \begin{center}
    \FigureFile(126mm,90.72mm){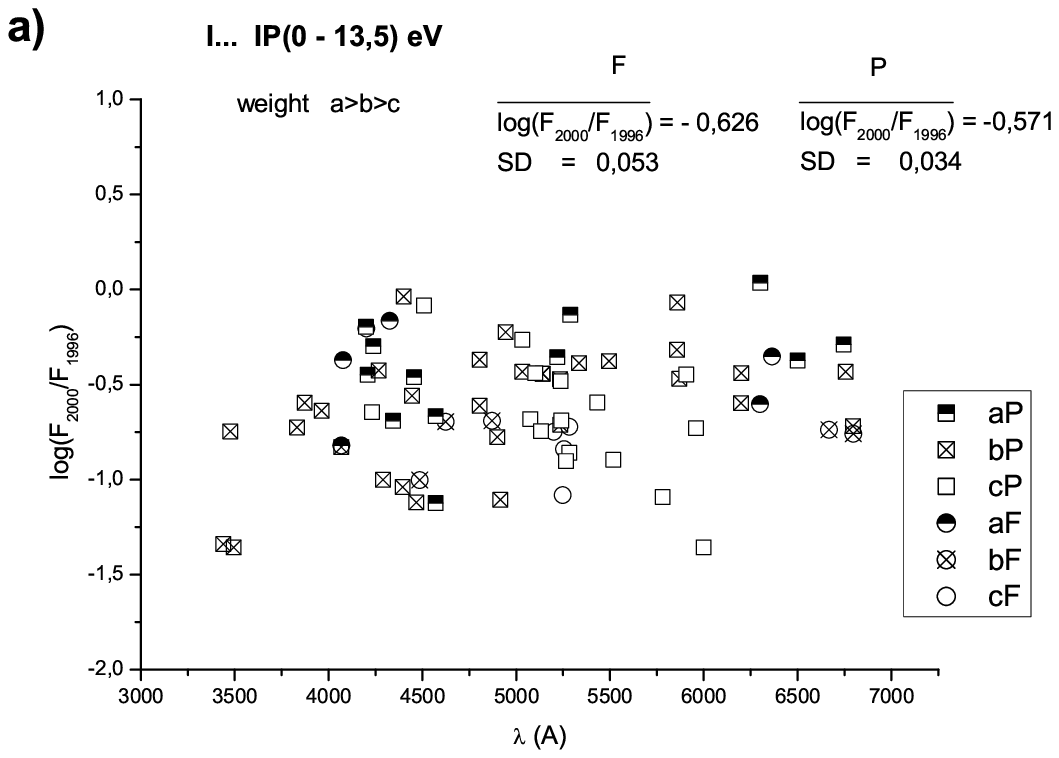}
  \end{center}
  \begin{center}
    \FigureFile(126mm,90.72mm){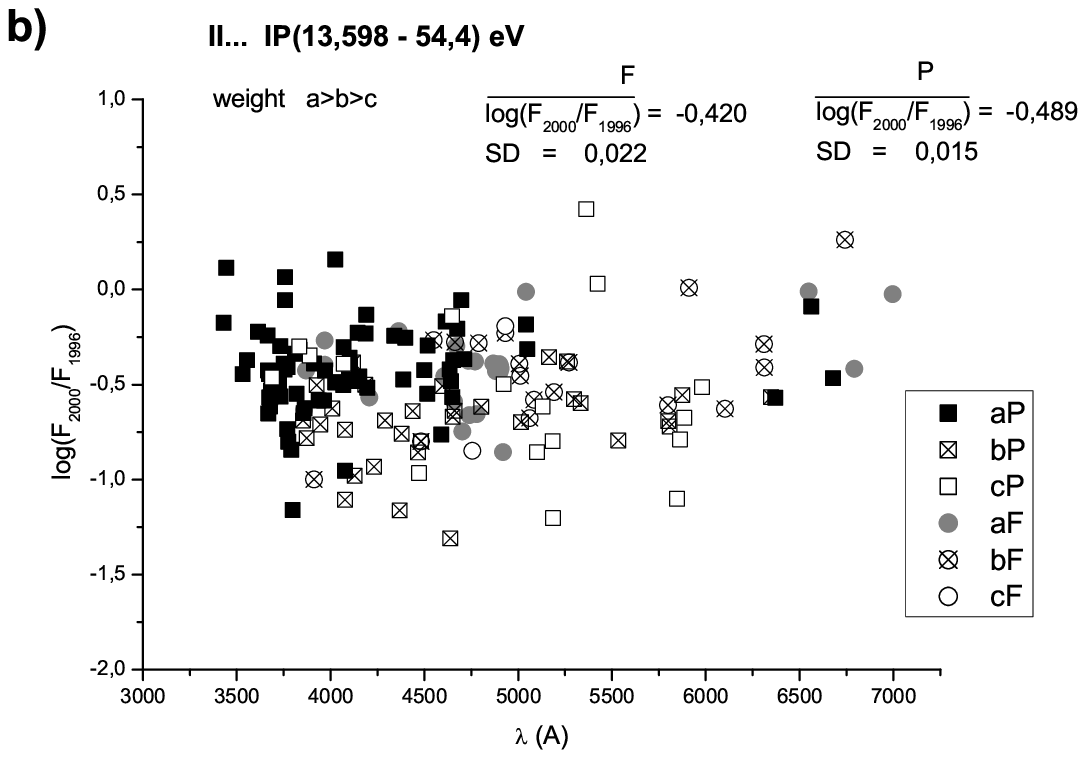}

  \end{center}
  \caption{$log(F_{2000}/F_{1996})$ vs. wavelength for ions other than FeII in the
potential range a) 1-13,5 eV, b) 13,598-54,4 eV, c) 54,416-77,472
eV, d) more than 77,5 eV. The mean values of
$log(F_{2000}/F_{1996})$ and standard deviations SD of the means for
the permitted and forbidden lines are given in the figure.
    $F_{2000}$ and $F_{1996}$ are emission line fluxes from the spectra
      taken 2000 and 1996 respectively. Permitted (P) lines are marked with rectangular and forbidden (F)
lines with circular symbols. "a" is for lines with statistical
weight 3, "b" for lines with statistical weight 2 and "c" for lines
with statistical weight 1.
      }\label{fig:prv}
\end{figure}

The behaviour of log flux ratios of the non-Fe+ lines from the 2000
and 1996 spectra is here examined. The plots of
$log(F_{2000}/F_{1996})$ against wavelength with the means and the
standard deviation of the means for different ranges of ionization
   potential determined for the permitted and forbidden
lines separately are shown in Figs.~\ref{fig:prv}a, 1b, 1c, and 1d.
  In the determination of the means and standard deviations, weights have
been given to the lines, depending to which of our three groups,
each line belongs. A plot of the
   mean log fadings against ionization potential is shown in Fig.~\ref{fig:dva}.
   The choice of the most appropriate
ionization potential for the contributing ions depends on the
mechanism of excitation. In the case of the forbidden lines of ions
and permitted metallic lines from singly ionized atoms (except for
the high excited lines of AlII and MgII), presumably excited from
the ground level by collisions, it is justified to adopt the
ionization potential of the previous ionization stage. For neutral
atoms a value of zero was assigned to the corresponding ionization
potential. Excitation from the ground level was assumed also for the
lines excited by the Bowen mechanism i.e. OIII 3430, 3444, 3754,
3757, 3774, 3791, 3810, the others not being excited in this way (
\cite{E08}). We have assumed that the Bowen mechanism is not
important for NIII, as found by Eriksson et al ( \cite{E05}), but we
must note that Selvelli et al ( \cite{Se07}) come to the opposite
conclusion, taking into account what happens when the optical
thickness is very large. We can also assume excitation from the
ground level for all lines arising from levels far from the
ionization limit not due to hydrogen or helium. Note that this
condition has been chosen quite arbitrarily. On the other hand, all
HI, HeI and HeII lines, as well as lines arising from levels closer
to the ionization level, have been taken to be excited by
recombination from the following stage of ionization, and the
ionization potential of the ion which produces the lines studied,
was adopted.

\begin{figure}
  \begin{center}
    \FigureFile(126mm,90.72mm){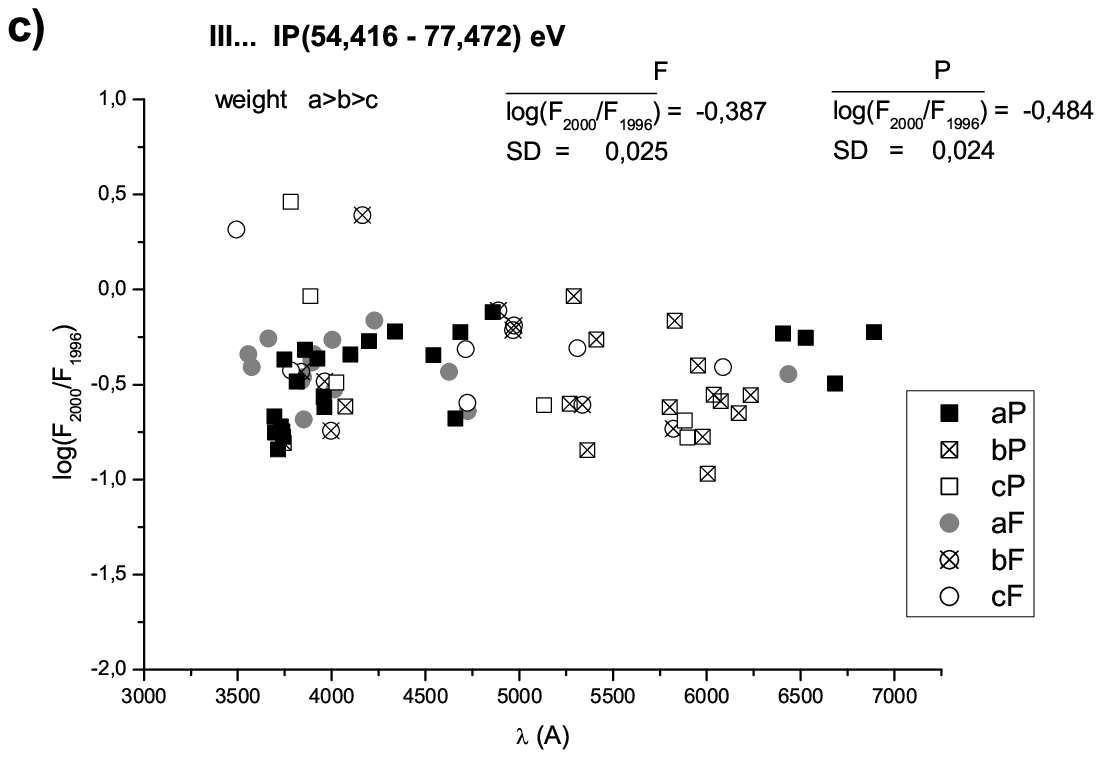}
  \end{center}
  \begin{center}
    \FigureFile(126mm,90.72mm){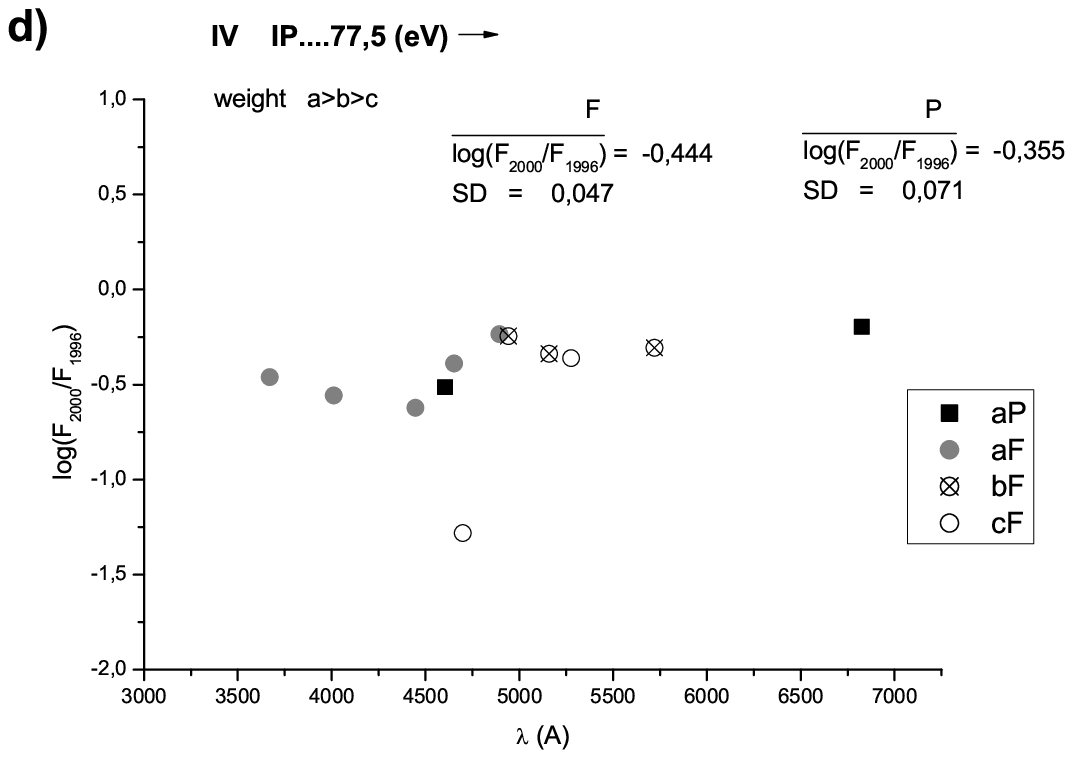}

  \end{center}
  \begin{center}
  \footnotesize{\textbf{Fig.~\ref{fig:prv}.} (continued)}
  \end{center}
\end{figure}

\newpage

The new additional evidence for fading of emision line fluxes in
2000 during a dust obscuration event with respect to the fluxes in
1996, shown in Figs. 1a, 1b, 1c, 1d and 2, might be considered
surprising, in view of the fact that (\cite{Se07}) found no
indication for wavelength dependent absorption in the optical. It is
therefore useful to check the reliability of the data used by us. We
previously found no evidence for fading at an earlier time
(\cite{KK02}) by comparing the measurements of \cite{Cr99} with
those of \cite{Mc97}. It is imaginable that the apparent fading
could be an artifact due to a non-linearity of the 2000 calibration
spectrum. To eliminate this as a possible explanation of our result,
we have looked for a dependence of fading on line center flux, the
latter being indicated by a ratio of total line flux in 2000
($F_{2000}$) to the corresponding FWHM. Such a test is best
undertaken for both the permitted Fe II and the forbidden [FeII]
lines, where lines not fitting the Self-Absorption Curve relation
between a function of flux and oscillator strength were eliminated
as badly identified. That method, previously used by us
(\cite{KK02}) was applied in a different way to another star by
Muratorio et al (\cite{Mu06}). The result of the test for
non-linearity is shown in the graph of Fig.~\ref{fig:tri} where a
small tendency for log $F_{2000}/F_{1996}$ to increase as a function
of log $F_{2000}/FWHM$ has been found. A calculation of the
correlation suggests an increase of 0.07 between values of log
F2000/FWHM of  -4 and -2 for the permitted lines and an increase of
0.19 for the forbidden lines over the same range. Even if the effect
is really due to non-linearity, it is clear that the mean fadings
for the two groups of emission lines of ionized iron (\cite{KF06}),
as well as those found in the present work, are considerably larger.

\begin{figure}
  \begin{center}
    \FigureFile(126mm,90.72mm){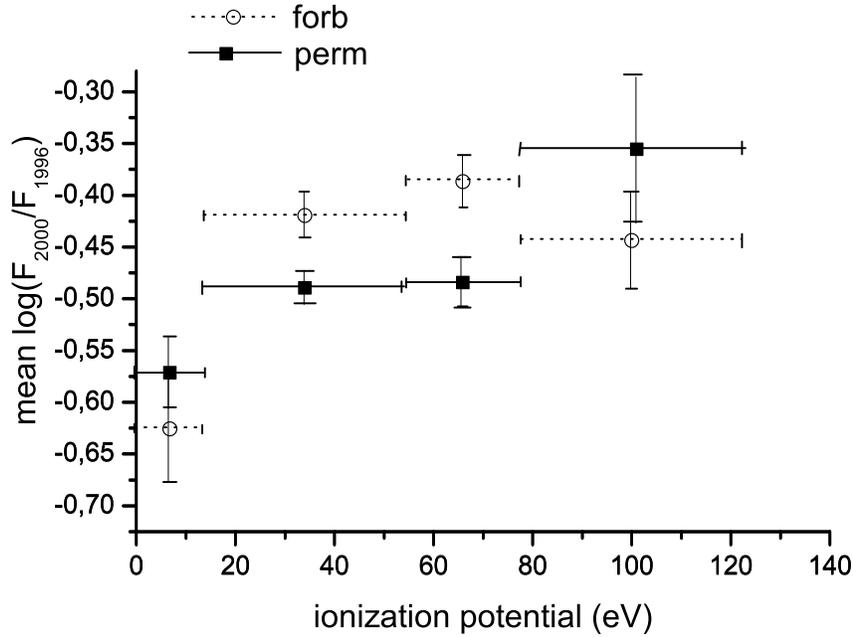}
  \end{center}
    \caption{ $log(F_{2000}/F_{1996})$ vs. ionization potential for ions other than FeII averaged
    over four ionization potential ranges: 1-13,5 eV; 13,598-54,4 eV;
    54,416-77,472 eV; more than 77,5 eV. The horizontal bars indicate ionization
potential ranges.
      }\label{fig:dva}
\end{figure}

\begin{figure}
  \begin{center}
    \FigureFile(126mm,90.72mm){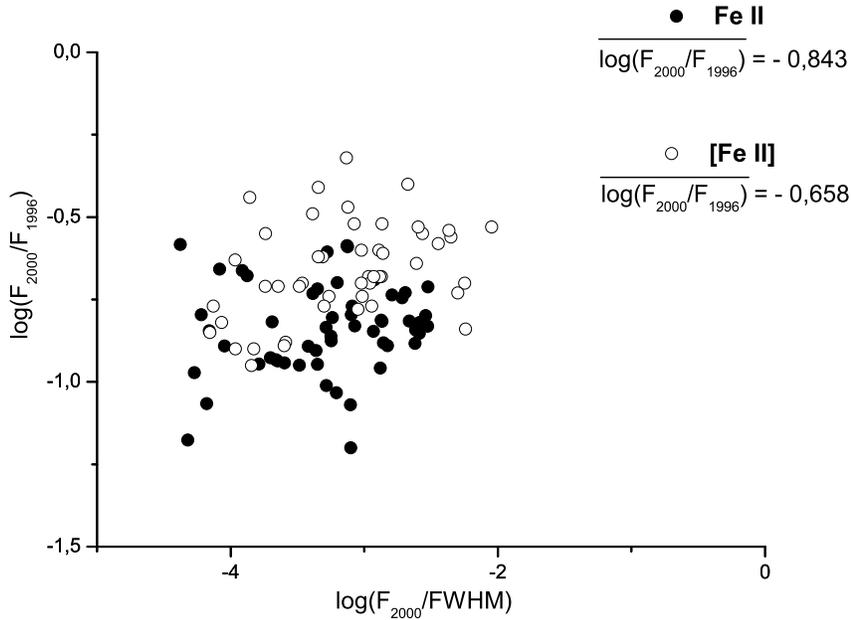}
  \end{center}
  \caption{$log(F_{2000}/F_{1996})$ vs. log $F_{2000}/FWHM$ for the Fe$^+$ lines.
  The values of log $F_{2000}/FWHM$ in the abscissa give a measure of the
  line peak fluxes. $F_{2000}$ and $F_{1996}$ are emission line fluxes from the spectra
      taken 2000 and 1996 respectively while FWHM refers to lines
      from the spectra taken in 2000.
      }\label{fig:tri}
\end{figure}

\newpage

\section{Discussion}
 We only have emission line fluxes for two dates, one inside and one outside
an obscuration event, so any interpretation is provisional. More
observations are needed to confirm or invalidate what is seen here.

The graphs of log flux ratio against wavelength
(Fig.~\ref{fig:prv}) do not show any very clear sign of any
correlation between these quantities. Within the errors the fits
indicate an absence of a slope.

We have however studied the variation of the means and the standard
deviations in the different ranges of ionization potential which are
indicated by horizontal bars (Fig.~\ref{fig:dva}). The standard
deviations are larger than for Fe$^+$ studied by itself (
\cite{KF06}) and unlike for Fe$^+$ show no significant difference
between permitted and forbidden lines. Whether this, like the larger
standard deviations, found here, is at least due to a certain extent
to different ions and atoms being studied together in the present
investigation, is not clear. The different permitted line fluxes are
of course affected by radiative transfer effects.

All the mean log ratios are less than zero, so fading occurs for
both permitted and forbidden lines in all ranges of ionization
potential. It appears to be however largest in the lowest range of
ionization potential.

This might indicate that the optically thick clouds whose existence
in the Fe$^+$ regions was first suggested by Kotnik-Karuza et al
(\cite{KF06}) are also present in regions of formation of higher
ionization lines. The clouds absorbing emission lines in the optical
and ultraviolet ranges are further out from the cool component than
the dust producing infrared absorption.

The latter region then should have a smaller covering factor.
Selvelli et al (\cite{Se07}) however found apparently contradictory
conclusions from the same HST/STIS data of October 2000, which
calibrate earlier spectra obtained with the Anglo-Australian
telescope, studied by us, The relative fluxes of the HeII Fowler
series indicated a fairly low interstellar reddening of E(B-V) near
0.00, agreeing with what is expected from the interstellar hydrogen
column density. However grey absorption by optically thick clouds
would not be detectable by a procedure  which only compared fluxes
of lines at different wavelengths.

We have determined ${\Delta}$E(B-V), i.e. the change of E(B-V) in
2000 with respect to its value in 1996 separately for the forbidden
and permitted lines of Fe$^+$ and for other ions in different
ionization potential ranges as shown in Table~\ref{tab:tri}.

\newpage
\setcounter{table}{2}
\begin{table}
\caption{$\Delta E(B-V)$ for Fe$^{+}$ lines and for lines other than
Fe$^{+}$ in four ionization potential ranges as defined in Fig.1.}
\label{tab:tri} \centering
\begin{tabular}{cc}
\hline &$\Delta E(B-V)$\\
\hline
IP I forb  & $-0.14 \pm 0.21$ \\
IP I perm  & $ 0.30 \pm 0.14$ \\
IP II forb  & $ 0.39 \pm 0.12$ \\
IP II perm  & $-0.10 \pm 0.09$ \\
IP III forb  & $ 0.04 \pm 0.12$ \\
IP III perm  & $-0.07 \pm 0.11$\\
IP IV forb  & $ 0.41 \pm 0.48$ \\
IP IV perm  & $ 0.40 $ \\
Fe$^{+}$ forb  & $ 0.10 \pm 0.10$ \\
Fe$^{+}$ perm  & $ 0.09 \pm 0.07$ \\

\hline
\end{tabular}
\end{table}

In view of the errors, the values of ${\Delta}$E(B-V) which are in
some cases negative are not significant.
 We may note that an
increase of the photospheric temperature of the hot component, would
lead to positive log flux ratios for the highest ionization lines,
formed nearer that component, unlike what we have found. An
alternative possibly more acceptable interpretation to that of the
last paragraph, would require the hot component to have moved to the
cooling part of its track in the temperature luminosity diagram,
following its increasing temperature in earlier decades, shown by
the appearance with time of lines from ions with increasing
ionization potential.

\begin{figure}
  \begin{center}
    \FigureFile(126mm,90.72mm){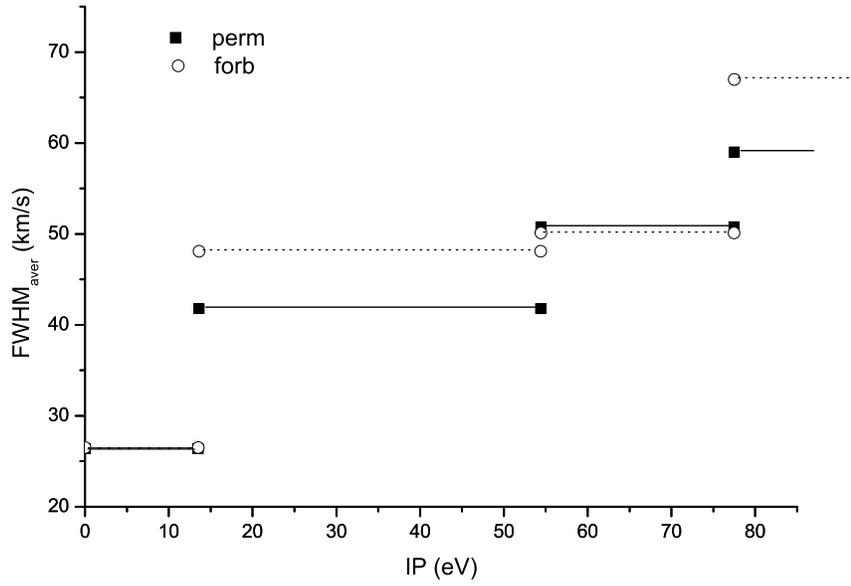}
  \end{center}
    \caption{ FWHMs of the permitted and forbidden lines from the spectra taken in 2000 averaged
    over four ionization potential ranges: 1-13,5 eV; 13,598-54,4 eV; 54,416-77,472 eV; more than 77,5 eV.
    Note that the permitted and the forbidden lines overlap in the lowest ionization potential range.
     In the averaging procedure over different ionization potential ranges each ion was weighted by the number of its lines.
The horizontal bars indicate ionization potential ranges.
 }\label{fig:cetri}
\end{figure}

\begin{figure}
  \begin{center}
    \FigureFile(126mm,90.72mm){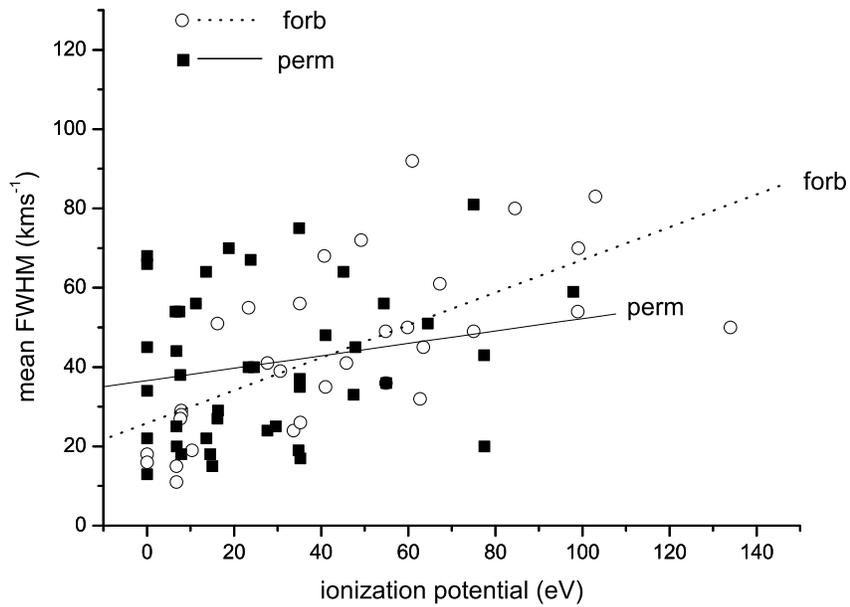}
  \end{center}
    \caption{Mean FWHMs of lines belonging to individual ions from
      the spectra taken in 2000 as a function of ionization
      potential
      }\label{fig:pet}
\end{figure}

The 2000 spectrum of RR Tel still shows the previously known effect
(see \cite{T77}) that lines from more highly ionized atoms tend to
be wider. This can clearly be seen in Fig.~\ref{fig:cetri}, where
the mean FWHM is plotted against ionization potential in the same
ranges of ionization potential as in Fig.~\ref{fig:dva}. It is also
shown in Fig.~\ref{fig:pet}, where the mean FWHM for each ion from
Table~\ref{tab:dva} is plotted against its ionization potential.
Fig.~\ref{fig:pet} shows the mean FWHM of individual ions.
Fig.~\ref{fig:cetri} shows averages
   over different ionization potential ranges indicated by horizontal
   bars. The contribution of each ion to each
   ionization potential range has been weighted by the number of its measured
   lines.

Two discordant lines have been left out from the plot: the 6742
[MnIII] and the 6548[NII] line. The 6742 [MnIII] line is the only
identified line of this ion in the 2000 spectrum but not identified
before. Its radial velocity shift is relatively large with respect
to other lines in its neighbourhood. Its flux is moreover quite low
and the large FWHM suggests that it is very weak. The 6548[NII] line
is the only line of multiplet 1F identified in our spectrum, in the
1996 spectrum it was declared as a blend and had not been identified
in the preceeding 1992 spectrum. This line, accompanied by the three
times stronger 6583[NII] line of the same multiplet, is a common
feature of normal nebulae. The absence of the stronger 6583[NII]
line in our 2000 spectrum indicates that the identification of the
weaker 6548[NII] line is probably wrong and shows that the emission
line regions of RR Tel are not like such nebulae.

According to Friedjung (\cite{F66}), the FWHMs of the RR Tel
emission lines of a given ionization stage tended to decrease with
time. Data given in that paper for HeII (-, 104 km s$^{-1}$), [Fe V]
(92 km s$^{-1}$, 81 km s$^{-1}$), [FeVI] (182 km s$^{-1}$, - ) and
[FeVII] (169 km s$^{-1}$, 182 km s$^{-1}$) refer to the spectra
taken in 1960 and 1965 respectively. That data from photographic
spectra at wavelengths below 5000 \AA\, with a quite approximate
correction for instrumental broadening are of lower quality than
those of the present paper and it is not easy to make a detailed
comparison. However, by comparing the non-instrumental broadening
corrected $FWHM_{2000}$ means of Table~\ref{tab:dva} with the
previously found values from 1960 and 1965 spectra, we can see that
the mean FWHM decrease of the wider lines of highly ionized atoms
seems to have continued.

The radii of the hot component photosphere of RR Tel at different
dates, estimated by M\"urset and Nussbaumer (\cite{MN94}), suggest
that these velocities were considerably less than the hot component
escape velocities and therefore of a hot component wind. Let us in
addition note that, as mentioned by Friedjung (\cite{F66}), the
Balmer lines near the beginning of the Balmer series, still tended
to be wider in 2000, this being especially true for H$\alpha$ .

\
\
\
\newpage

\textbf{Acknowledgements}. We wish to thank Mattias Eriksson for
providing us useful information on the excitation mechanism for
certain OIII lines. Our thanks are also due to the referee R. Viotti
for his inspiring remarks and suggestions which helped us to improve
this paper a lot.

\end{document}


\begin{longtable}{*{12}{l}}
\caption{List of emission lines from the spectra of RR Tel taken in
2000 and 1996} \label{tab:first}\hline \hline
$\lambda$& Ion & mult & IP& $F_{2}$ & FWHM & $F_{1}$ & log($F_{2}$/$F_{1}$)& exc & w\\
(\AA)& & & (eV)&& (km/s) & & & & \\
\hline
\endhead
\hline

\endfoot
\hline \multicolumn{12}{l}{$F_{1}$, $F_{2}$... line fluxes from the
spectra taken in 1996 and 2000 respectively, given in units of
10$^{-12}$ergs $^{-1}$cm$^{-2}$

} \\
\multicolumn{12}{l}{mult ... multiplet number} \\
 \multicolumn{12}{l}{IP ... ionization potential of the corresponding ion}\\
\multicolumn{12}{l}{FWHM ... of lines from the spectrum taken in 2000}\\
\multicolumn{12}{l}{exc ... excitation mechanism: g - collisions
from the ground state, B - Bowen mechanism,
 r - recombination }\\
\multicolumn{12}{l}{w ... statistical weight: a$=$3, b$=$2, c$=$1,
x...means that the line is beyond the limits of the studied wavelength range}\\

\hline
\endlastfoot

3183.61 &   MnII    &       &   7.434   &   4.7E-02 &   19  &   1.1E-01 &   -0.37   &   g   &   x   \\
3185.72 &   OIV &   7   &   77.413  &   2.1E-02 &   19  &   4.6E-02 &   -0.35   &   r   &   x   \\
3187.74 &   HeI &   3   &   24.587  &   1.2E-01 &   53  &   3.8E-02 &   0.50    &   r   &   x   \\
3193.10 &   SiII    &       &   16.346  &   1.7E-02 &       &   5.1E-02 &   -0.48   &   r   &   x   \\
3196.07 &   FeII    &   7   &   7.902   &   7.6E-02 &   19  &   9.7E-02 &   -0.11   &   g   &   x   \\
3203.10 &   HeII    &   1   &   54.416  &   3.82    &   41  &   5.01    &   -0.12   &   r   &   x   \\
3213.31 &   FeII    &   6   &   7.902   &   6.0E-02 &   21  &   8.2E-02 &   -0.14   &   g   &   x   \\
3220.84 &   FeII    &   106 &   7.902   &   2.0E-01 &   57  &   1.6E-01 &   0.09    &   g   &   x   \\
\end{longtable}